# Investigating quarkonium collectivity in heavy-ion collisions

Victor Valencia Torres for the ALICE collaboration
*SUBATECH, IMT Atlantique, Nantes Université, CNRS/IN2P3, Nantes, France*

Quarkonia have long been identified as ideal probes for understanding the quark-gluon plasma (QGP). Heavy quarks are produced in the early stage of the heavy-ion collisions, therefore they experience the evolution of the medium produced, providing an important tool to investigate the properties of the QGP. In particular, the positive elliptic flow measured at the LHC is interpreted as a signature of the heavy-quark thermalization and recombination in the QGP. This is reflected in the azimuthal anisotropies in the final particle production. A better understanding of heavy-quark energy loss, quarkonium dissociation, and production mechanism can therefore be obtained with the elliptic flow observable. We present new flow results of quarkonia in Pb–Pb collisions at $\sqrt{s_{\text{NN}}} = 5.36$ TeV at forward rapidity carried out with the ALICE detector.

## Introduction

The production of heavy quarks originates in the initial stages of hadronic collisions via hard-scattering processes and can be described by perturbative Quantum Chromodynamics (pQCD). In heavy-ion collisions, such as Pb–Pb, heavy quarks traverse the entire evolution of the quark-gluon plasma (QGP), making them powerful probes of the medium's properties. Studying their possible thermalization within the QGP provides valuable insights into the nature of the strongly interacting matter. By studying quarkonium production, one can investigate whether heavy quarks participate to the collective motion of the medium. Flow observables, such as the elliptic flow coefficient $v_2$, are key tools for characterizing this behavior[1].

The ALICE apparatus[2] measures inclusive charmonia at forward rapidity ($2.5 < y < 4$) through the dimuon decay channel. At midrapidity, prompt and non-prompt heavy-flavor hadrons (with non-prompt originating from the decay of longer-lived hadrons) can be well distinguished. The key detectors at midrapidity include the Inner Tracking System (ITS), which is essential for tracking, vertexing, and measuring charged-particle multiplicity, and the Time Projection Chamber (TPC), which is utilized for tracking and particle identification by measuring specific energy loss. At forward rapidity, the muon spectrometer, consisting of a system of absorbers, a dipole magnet, and tracking and MID stations, is used for reconstructing and identifying muon tracks. The FIT detectors (FT0A and FT0C), measures particle multiplicity, centrality and event-plane angle for the flow measurements. Additionally, inclusive hadrons at both forward and midrapidity can be measured down to zero $p_{\text{T}}$.

## Results

We present for the first time the $v_2$ measurements of quarkonia in Pb–Pb collisions at $\sqrt{s_{\text{NN}}} = 5.36$ TeV with the data collected in 2023. The significantly larger Run 3 dataset compared to Run 1 and 2 enable more precise flow measurements. Different flow methods are performed, the event-plane (EP), the scalar-product (SP) and for the first time a new additional method: the multi-particle cumulant. The extraction of the $v_2$ is performed with an event-mixing technique to reproduce the combinatorial background of Pb-Pb collisions.

The $J/\psi$ $v_2\{SP\}$ results as a function of the event centrality in two $p_T$ ranges are reported in Fig. 1.

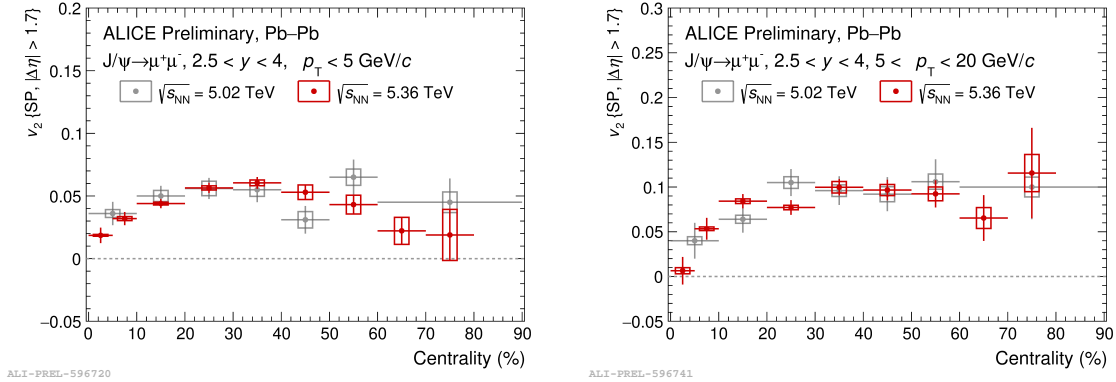

Figure 1 – Comparison of $v_2^{J/\psi}\{SP\}$ for $2.5 < y < 4$ between Run 3 (red) and Run 2 (grey) as a function of centrality for $p_T < 5$ GeV/c (left panel) and $5 < p_T < 20$ GeV/c (right panel). Error bars represent statistical uncertainties. Empty boxes indicate systematic uncertainties.

The centrality dependence of $v_2$ (red points) measured in Run 3 is consistent with the Run 2 measurements (grey points) while providing an improved granularity as a function of centrality. In the low-$p_T$ region, where regeneration takes place[3], a smooth centrality dependence is observed, with a peak around 30% centrality. At high $p_T$, $v_2$ is near zero in the most central bin, increasing with centrality, in line with expectations from path-length dependent energy-loss mechanisms.

The $J/\psi$ $v_2$ results as a function of $p_T$ in the centrality classes 0–10%, 10–30%, 30–50%, and 50–80% are reported in Fig. 2.

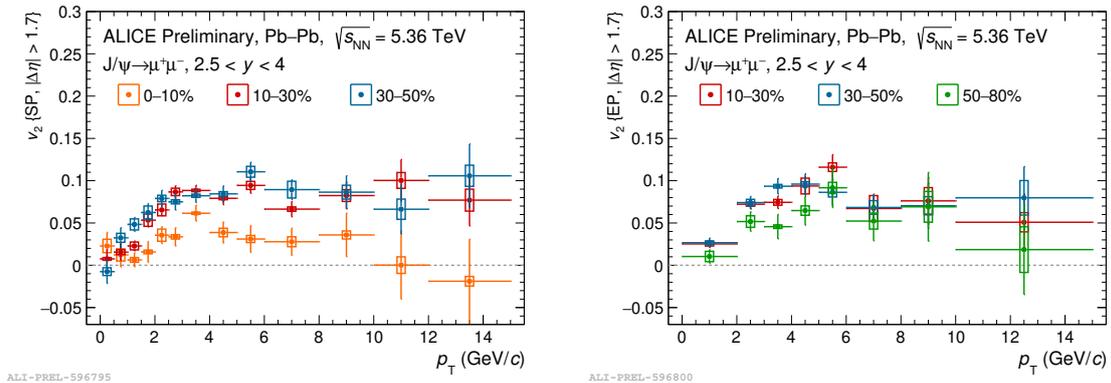

Figure 2 – $v_2^{J/\psi}\{SP\}$ as a function of $p_T$ in the centrality classes 0–10%, 10–30%, and 30–50%, measured using the scalar product (left-panel). $v_2^{J/\psi}\{EP\}$ as a function of $p_T$ in the centrality classes 10–30%, 30–50%, and 50–80%, measured using the event plane (right-panel).

The results from the SP (left-panel) and EP (right-panel) methods are in agreement. A positive $v_2$ is observed over a broad $p_T$ range (from 0 up to 10 GeV/c) for the different centrality classes. Furthermore, there is a hint of centrality dependence. In non-central collisions, a positive $v_2$ is expected from the initial eccentricity[1] ($v_2 \sim \kappa_2\epsilon_2$). At low $p_T$, the significant $v_2$ is understood by the regeneration scenario. At higher $p_T$, where path-length dependent energy-loss effects are expected, $v_2$ seems to be flat with slightly positive values. However, due to significant statistical uncertainties beyond 10 GeV/c, no definitive conclusions can be drawn in this region.

In the 0–10% centrality class, $v_2$ shows a weaker dependence on $p_T$, with lower peak values and a decreasing trend at high $p_T$. The first measurement within the 50–80% centrality range was also performed in the event plane and is shown on Fig. 2 right.

We also employed a new multi-particle cumulant technique[4] to extract the flow of $J/\psi$, based on correlations between charged particles at midrapidity and dimuons at forward rapidity. For the first time, we present measurements of the $J/\psi$ flow coefficients: $v_2^{J/\psi}\{m\}$, for $m$-particle correlations ($m = 2$ and $m = 4$) in Fig. 3.

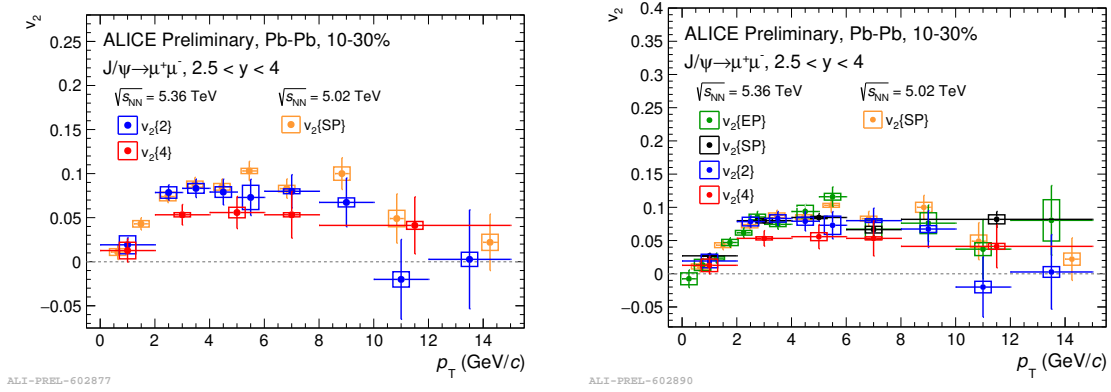

Figure 3 – The $J/\psi$ $v_2$ results for $2.5 < y < 4$ as a function of $p_T$ in the 10–30% centrality class. Left panel: comparison between $v_2\{2\}$ and $v_2\{4\}$ measurements from Run 3 and the $v_2\{SP\}$ from Run 2. Right panel: comparison of all Run 3 flow methods (EP, SP, and cumulants) with the $v_2\{SP\}$ result from Run 2.

There is an agreement between $v_2\{EP\}$, $v_2\{SP\}$ and $v_2\{2\}$ validating the robustness of the observed $J/\psi$ anisotropy. Assuming Gaussian fluctuations[5], it is expected that

$$v_2\{EP\}^2 \sim v_2\{SP\}^2 \sim v_2\{2\}^2 = \langle v_2 \rangle^2 + \sigma_v^2, \quad (1)$$

$$v_2\{4\} = \langle v_2 \rangle^2 - \sigma_v^2, \quad (2)$$

where $\langle v_2 \rangle$ is the average elliptic flow with respect to the reaction plane and $\sigma_v$ represents its event-by-event fluctuations. A significance of $2.7\sigma$ is observed for $v_2\{4\} < v_2\{SP\}$, indicating for the first time a strong sensitivity to $J/\psi$ flow fluctuations.

In addition, the measurement of the $\Upsilon(1S)$ elliptic flow in Pb-Pb collisions at $\sqrt{s_{NN}} = 5.36$ TeV is reported in Fig. 4.

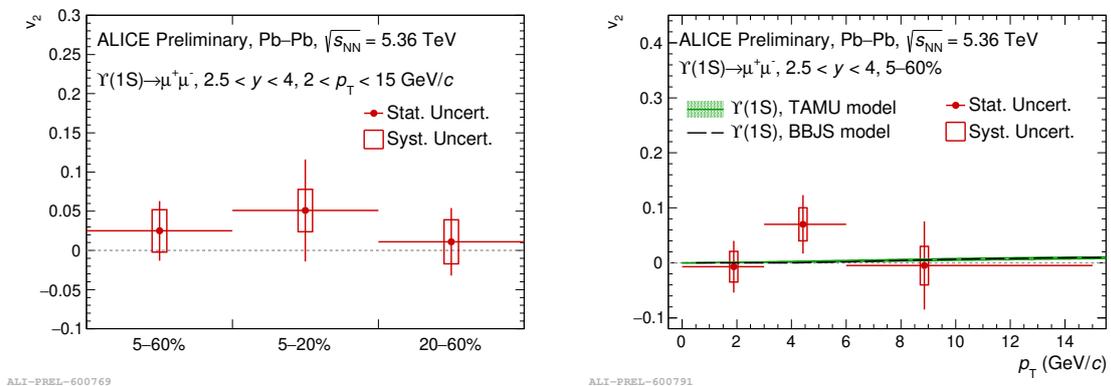

Figure 4 – $v_2\{EP\}$ of $\Upsilon(1S)$ for $2.5 < y < 4$ in Pb–Pb collisions at $\sqrt{s_{NN}} = 5.36$ TeV. Left panel: $v_2$ as a function of centrality. Right panel: $v_2$ as a function of $p_T$, including a comparison to models.

The $\Upsilon(1S)$ $v_2$ results are consistent with zero within the current uncertainties. In the low regeneration scenario, where the production of bottom quark pairs is scarce, a $v_2$ of $\Upsilon$ close to zero is expected. Notably, the TAMU model describes quarkonium suppression and regeneration within a hydrodynamically evolving QGP using rate equations[6]. It incorporates statistical approaches that account for medium-modified binding energies obtained from lattice QCD calculations. On the other hand, the BBJS model adopts a Schrödinger-Langevin dynamics framework[7], treating quarkonia as quantum states evolving under a complex in-medium potential. BBJS emphasizes wavefunction melting and stochastic suppression processes rather than explicit recombination. A qualitative agreement is observed between the experimental data and theoretical models. Nevertheless, the current uncertainties in the $\Upsilon(1S)$ $v_2$ measurement do not allow firm conclusions.

Finally, we also compare in Fig. 5 the new $v_2$ measurements of open heavy-flavour hadrons at midrapidity to the $v_2$ of quarkonia at forward rapidity.

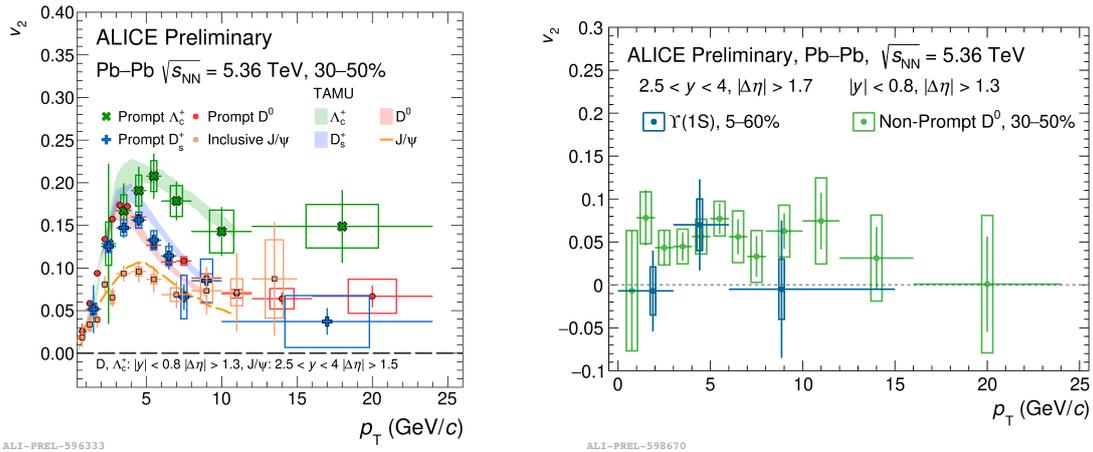

Figure 5 – Left panel: $v_2$ of prompt $\Lambda_c^+$, $D_s^+$ and $D^0$ at midrapidity vs. $J/\psi$ $v_2$ at forward rapidity. Right panel: $v_2$ of non-prompt $D^0$ at midrapidity vs $\Upsilon(1S)$ $v_2$ at forward rapidity.

The left panel of Fig. 5 shows a hierarchy of $v_2$ for hadrons containing a c quark at intermediate $p_T$: $v_2^{J/\psi} < v_2^{\text{prompt }D^0} \sim v_2^{\text{prompt }D_s^+} < v_2^{\text{prompt }\Lambda_c^+}$. This ordering supports the thermalization process of charm quark in QGP. The TAMU model reproduces the data qualitatively. Alternatively, other theoretical frameworks based on initial-state models, like CGC, predict that correlations among final-state particles could produce flow-like effects[8]. The right panel of Fig. 5 shows a hierarchy of $v_2$ for $\Upsilon(1S)$ and non-prompt $D^0$: $v_2^{\Upsilon(1S)} < v_2^{\text{non-prompt }D^0}$. These data add new constraints on the thermalization of the beauty quarks.